\documentclass[aip,preprint]{revtex4-1}
\usepackage{graphicx}

\usepackage{appendix}

\draft 

\begin{document}

\title{A Pseudo-Fermion Propagator Approach to the Fermion Sign Problem}

\author{Yunuo Xiong}
\email{xiongyunuo@hbpu.edu.cn}
\affiliation{Center for Fundamental Physics and School of Mathematics and Physics, Hubei Polytechnic University, Huangshi 435003, China}

\author{Hongwei Xiong}
\email{xionghongwei@hbpu.edu.cn}
\affiliation{Center for Fundamental Physics and School of Mathematics and Physics, Hubei Polytechnic University, Huangshi 435003, China}
\affiliation{Wilczek Quantum Center, Shanghai Jiao Tong University, Shanghai 200240, China}
\date{\today}

\begin{abstract}
In this work, within the framework of path integral Monte Carlo, we construct a pseudo-fermion propagator by replacing the original fermionic determinant with its absolute value. This modified propagator defines an auxiliary system free from the fermion sign problem, enabling efficient simulations of 
fermionic systems. We found that by shifting the pseudo-fermion energy based on the energy of a non-interacting fermion system, we can efficiently and reliably infer the energy of fermionic systems in various situations, from strong quantum degeneracy to weak quantum degeneracy. We have performed first-principles simulations of quantum dots confined in a two-dimensional harmonic potential and found excellent agreement with benchmark results provided by other established methods. We believe that this pseudo-fermion propagator framework opens up new possibilities for first-principles simulations of fermionic systems.
\end{abstract}

\maketitle

\section{Introduction}

Path integral Monte Carlo (PIMC) \cite{CeperleyRMP,CeperleyBook,Tuckerman,Fosdick,Jordan,barker,Morita,Burov1,Burov2} provides a powerful first-principles method for simulating identical bosons as well as boltzmannons. In this method, there is no need to assume any physical properties of the quantum system in advance, and, in principle, it enables exact simulations of the thermodynamic properties of large-scale quantum systems.
Mathematically, in PIMC, the quantum system is mapped to a large number of classical beads, and the partition function becomes a high-dimensional integral of the form:
\begin{equation}
Z = \int dx_1 \cdots dx_K \, e^{-f(x_1, \cdots, x_K)}.
\end{equation}
If we want to compute the energy, we construct an energy estimator $\epsilon(x_1, \cdots, x_K)$, so that the energy is given by:
\begin{equation}
E = \frac{\int dx_1 \cdots dx_K \, \epsilon(x_1, \cdots, x_K) e^{-f(x_1, \cdots, x_K)}}{\int dx_1 \cdots dx_K \, e^{-f(x_1, \cdots, x_K)}}.
\end{equation}
If $e^{-f(x_1, \cdots, x_K)}$ is positive, we can perform an extensive importance sampling on this complicated function. In this case, the energy can be expressed as:
\begin{equation}
E = \frac{\sum_j \epsilon(j)}{\sum_j}.
\end{equation}

Unfortunately, for fermionic systems, due to the antisymmetry of the identical fermion wave function under particle exchange, $e^{-f(x_1, \cdots, x_K)}$ is positive in some regions and negative in others, when the Trotter decomposition is employed, resulting in a large number of imaginary time slices. As a result, importance sampling over the entire domain becomes impossible. This is the so-called fermion sign problem \cite{Loh,ceperley,troyer,ZhangS, diagrammatic,Booth,Schoof,PB,Groth,Malone,Dornheim,Alex,Hou,WDM,HirshbergFermi,DornheimFermi}. 

Although the fermion sign problem seems notoriously difficult to solve, recent developments in the concept of fictitious identical particles \cite{XiongFSP,Xiong-xi,Dornheim1,Dornheim2,Dornheim3,Dornheim4,Dornheim5,Morresi1,Morresi2,Xiong-Hubbard,Yang,Dornheim6,Dornheim7,Bonitz-Review,RoadMap,Fan} inspire us to believe that, through general physical and mathematical analysis, there is still hope of overcoming the fermion sign problem in some important quantum systems, such as warm dense matter \cite{Dornheim1,Dornheim2,Dornheim3,Dornheim4,Dornheim5,Dornheim6,Dornheim7}, normal liquid $^3$He \cite{Morresi1} and Fermi-Hubbard model \cite{Fan}. In the fictitious identical particle approach, we introduce an additional variable $\xi$ into the function $e^{-f(x_1, \cdots, x_K)}$, such that the partition function becomes
\begin{equation}
Z(\xi) = \int dx_1 \cdots dx_K e^{-f(\xi, x_1, \cdots, x_K)}.
\end{equation}
Here, $\xi = 1, 0, -1$ correspond to bosons, boltzmannons, and fermions, respectively. By performing simulations in the bosonic sector ($\xi > 0$), it becomes possible to reliably and efficiently extrapolate to the thermodynamic properties of the fermionic system ($\xi = -1$). 

In the well-known fixed-node method \cite{fnm} and restricted path integral Monte Carlo \cite{RPIMC}, the fermion sign problem is addressed by restricting the integration domain of the high-dimensional function based on the general properties of fermions. In the fixed-node method, restricted path integral Monte Carlo, and the fictitious identical particle approach, once an appropriate quantum system is identified, the sign problem does not arise during the first-principles simulations. This allows these methods to demonstrate clear advantages in large-scale simulations of fermionic systems. The success of these methods provides important insight: by rethinking and reformulating the mathematical structure of the fermionic partition function from both physical and mathematical perspectives, new approaches may be developed.

In this work, we introduce pseudo-fermions, a novel class of fictitious particles distinct from the fictitious identical particles in Refs. \cite{XiongFSP,Xiong-xi}, aimed at addressing the fermion sign problem. Within the framework of PIMC, we consider the fermionic partition function expressed through the fermion propagator \cite{PB,latticeFermi,Miura,Takahashi,Lyubartsev1,Lyubartsev2,ChinF,Filinov1,Filinov2,PB2,PB3}. In this work, we propose a pseudo-fermion propagator to construct an auxiliary partition function, and we point out the new possibilities this auxiliary partition function offers for overcoming the fermion sign problem. 
For quantum dots confined in a two-dimensional harmonic potential, we find that the simulation results presented in this work are in excellent agreement with the benchmark results \cite{PB,Egger,Dornheim1} provided by other previous methods. In particular, we find that the pseudo-fermion method holds promise for efficiently and reliably inferring the energies of fermions across a wide range of conditions, from the ground state and strong quantum degeneracy to weak quantum degeneracy.

The structure of this paper is as follows: in Sec. \ref{FSP}, we introduce the fermion sign problem and the pseudo-fermion propagator. In Sec. \ref{fdp}, we describe the pseudo-fermion propagator and its general relationship to the energies of fermions. We analyze the reliability of using the pseudo-fermion propagator to simulate fermionic systems from a mathematical perspective. In Sec. \ref{results}, we present simulations of quantum dots in a two-dimensional harmonic potential and find excellent agreement with previous benchmark results. In Sec. \ref{summary}, we provide a brief summary and discussion.

\section{Fermion Sign Problem and Pseudo-Fermion Propagator}
\label{FSP}

\subsection{Fermion Sign Problem and Fermion Propagator}

We consider a system of $N$ spin-polarized fermions. The partition function is given by
\begin{equation}
Z_F(\beta, \lambda) = \mathrm{Tr}\left( e^{-\beta \hat{H}} \right).
\end{equation}
Here, $ \beta = 1 / (k_B T) $, where $ T $ is the temperature and $ k_B $ is the Boltzmann constant. The Hamiltonian operator consists of the kinetic energy operator  $ \hat{T} $, the potential energy operator  $ \hat{V}_{\mathrm{pot}} $, and the interaction energy operator between particles $ \hat{V}_{\mathrm{int}} $:
\begin{equation}
\hat{H} = \hat{T} + \hat{V}_{\mathrm{pot}} + \hat{V}_{\mathrm{int}}(\lambda).
\end{equation}
Here, $ \lambda $ represents the coupling strength of the interactions between particles.

In the absence of interactions between particles, for the part $ \hat{T} + \hat{V}_{\mathrm{pot}} $, we can always obtain the single-particle energy spectrum through numerical calculations and thus accurately compute thermodynamic properties such as the average energy at different temperatures. However, the presence of $\hat{V}_{\mathrm{int}}(\lambda) $ poses significant challenges.

We can write the partition function of the fermionic system as:
\begin{equation}
Z_F(\beta, \lambda) = \frac{1}{N!} \int d\textbf{R} \sum_P (-1)^{N_P} \left< P\textbf{R} \right| e^{-\beta \hat{H}} \left| \textbf{R} \right>.
\end{equation}
Here, \( \textbf{R}\equiv (\textbf{r}_1, \cdots, \textbf{r}_N)\) includes the coordinates of all \( N \) particles. \( P \) represents the permutation operator acting on the coordinates, and \( N_P \) denotes the minimal number of pairwise exchanges required to restore the original order of the coordinates under permutation \( P \). 
The presence of the factor \( (-1)^{N_P} \) in the fermionic partition function \( Z_F(\beta, \lambda) \) leads to the fermion sign problem \cite{Dornheim}, compared with the partition function of bosons.

We can express the partition function of the fermionic system as:
\begin{equation}
Z_F(\beta, \lambda) = \frac{1}{N!} \int d\textbf{R} \sum_P (-1)^{N_P} \left< P\textbf{R} \right| e^{-\Delta\tau \hat{H}} \cdots e^{-\Delta\tau \hat{H}} \left| \textbf{R} \right>.
\end{equation}
Here, \( \Delta \tau = \beta / M \), where \( M \) is the number of imaginary time slices. When \( \Delta \tau \) is small, we can apply the Trotter decomposition \cite{trotter} and insert appropriate identity operators over momentum and position to transform the above partition function into a high-dimensional integral.

When inserting the identity operators over positions, the most popular approach \cite{CeperleyRMP,CeperleyBook} is to insert the following identity operator for distinguishable particles:
\begin{equation}
\hat{I}_D = \int d\textbf{r}_1 \cdots d\textbf{r}_N \left| \textbf{r}_1 \cdots \textbf{r}_N \right> \left< \textbf{r}_1 \cdots \textbf{r}_N \right|.
\end{equation}
Apart from the above form of inserting identity operators, one can also insert the following operator \cite{Takahashi}:
\begin{equation}
\hat{I}_F = \frac{1}{N!} \sum_P (-1)^{N_P} \int d\textbf{r}_1 \cdots d\textbf{r}_N \left| \textbf{r}_1 \cdots \textbf{r}_N \right> \left< P\{ \textbf{r}_1 \cdots \textbf{r}_N \} \right|.
\end{equation}
In this case, we can define the fermion propagator \cite{PB,latticeFermi,Takahashi,Miura,Lyubartsev1,Lyubartsev2,ChinF,Filinov1,Filinov2,PB2,PB3} between two adjacent imaginary time slices as:
\begin{equation}
\rho_F(\textbf{R}^j, \textbf{R}^{j+1}) = \frac{1}{N!} \sum_P (-1)^{N_P} \left< P\textbf{R}^j \right| e^{-\Delta\tau \hat{H}} \left| \textbf{R}^{j+1} \right>.
\end{equation}
Here, the superscript \( j \) in \( \textbf{R}^j \) denotes the \( j \)-th imaginary time slice, where \( j = 1, 2, \cdots, M \). In addition, we impose the condition \( \textbf{R}^{M+1} \equiv \textbf{R}^1 \).

Using the Trotter decomposition \cite{trotter}, we have
\begin{equation}
\rho_F(\textbf{R}^j, \textbf{R}^{j+1}) = D_{\mathrm{free}}(\textbf{R}^j, \textbf{R}^{j+1}; \Delta\tau) e^{-\frac{\Delta\tau}{2} ( V_{\mathrm{pot}}(\textbf{R}^j) + V_{\mathrm{pot}}(\textbf{R}^{j+1}) )} e^{-\frac{\Delta\tau}{2} ( V_{\mathrm{int}}(\textbf{R}^j) + V_{\mathrm{int}}(\textbf{R}^{j+1}) ) }.
\end{equation}
Here,
\begin{equation}
D_{\mathrm{free}}(\textbf{R}^j, \textbf{R}^{j+1}; \Delta\tau) = \frac{1}{N!}\sum_P (-1)^{N_P} \left< P\textbf{R}^j \right| e^{-\Delta\tau \hat{T}} \left| \textbf{R}^{j+1} \right>
\end{equation}
is the antisymmetric free-fermion propagator. After inserting the identity operator over momentum, \( D_{\mathrm{free}}(\textbf{R}^j, \textbf{R}^{j+1}; \Delta\tau) \) becomes a determinant:
\begin{equation}
D_{\mathrm{free}}(\textbf{R}^j, \textbf{R}^{j+1}; \Delta\tau) = \frac{1}{N!}\det\left(\frac{1}{(2\pi\Delta\tau)^{d/2}}\exp\left(-\frac{1}{2\Delta\tau}\left(\textbf {R}^j_{l}-\textbf {R}^{j+1}_m\right)^2\right)\right).
\end{equation}
Here, $d$ denotes the spatial dimension of the fermionic system. In $\textbf{R}^j_l$, the superscript $j$ indicates the $j$-th imaginary time slice, and the subscript $l$ labels the $l$-th particle. It is worth noting that the determinant \( D_{\mathrm{free}}(\textbf{R}^j, \textbf{R}^{j+1}; \Delta\tau) \) can be either positive or negative, and its sign does not depend on the potential energy or the interactions between particles.

\begin{figure}[htbp]
\begin{center}
\includegraphics[width=0.8\textwidth]{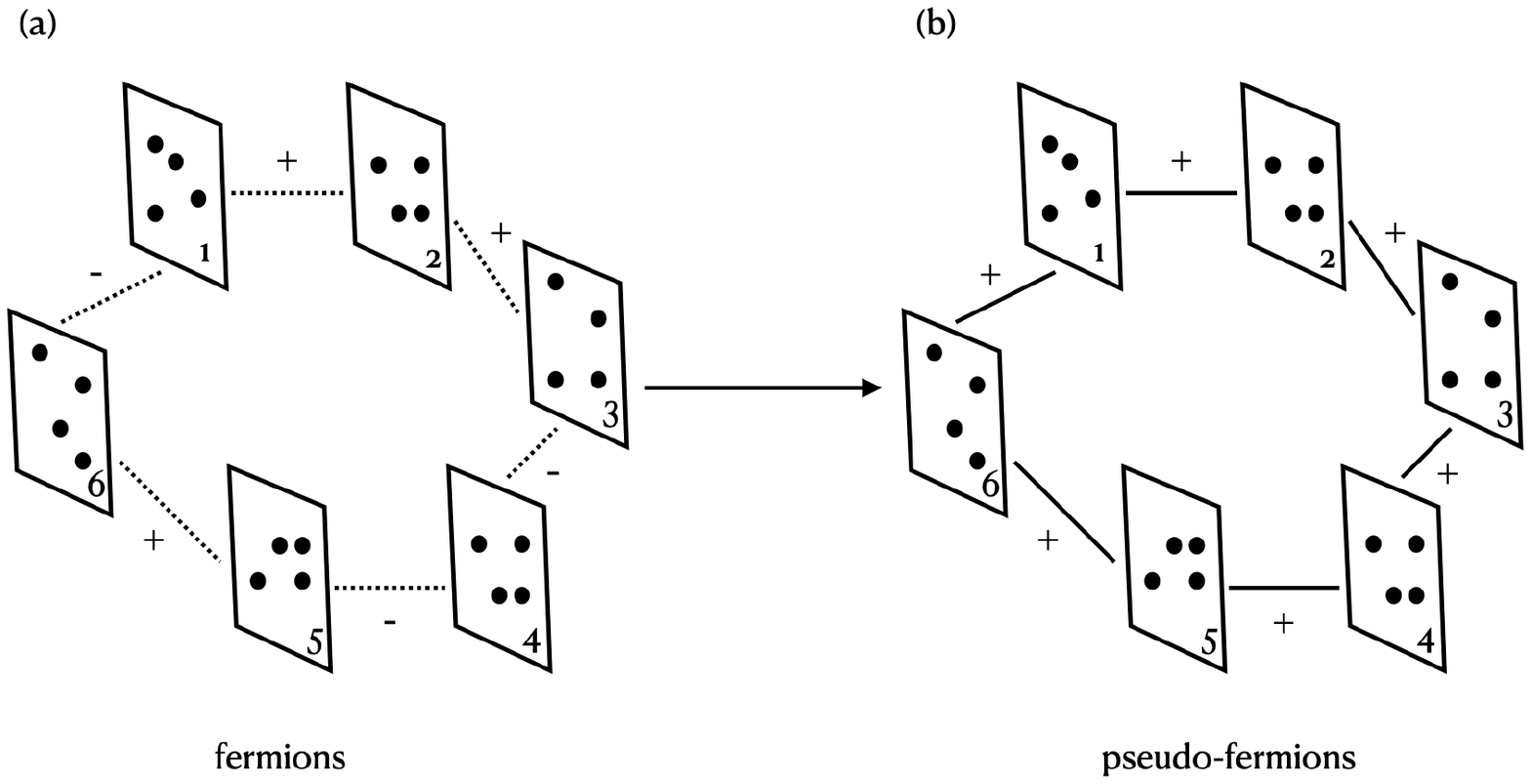}
\caption{\label{Figure1} (a) shows the positions of 4 fermions across 6 imaginary time slices. The dashed lines between two adjacent imaginary time slices represent fermion propagators. The determinant within each fermion propagator can be positive or negative, which we denote by "+" or "-" next to the dashed lines. For the case shown in (a), since there is an odd number of "-", the overall sign is negative. (b) shows the case of the pseudo-fermion propagator. The pseudo-fermion propagators are represented by solid lines, where the determinant in each fermion propagator is replaced by its absolute value. Therefore, for pseudo-fermions, the sign problem does not exist during the PIMC simulation.}
\end{center}
\end{figure}

Using the free-fermion propagator, the partition function of the fermionic system is given by
\begin{equation}
Z_F(\beta, \lambda) = \int \prod_{j=1}^M D_{\mathrm{free}}(\textbf{R}^j, \textbf{R}^{j+1}; \Delta\tau)
e^{-\Delta\tau V_{\mathrm{pot}}(\textbf{R}^j)} e^{-\Delta\tau V_{\mathrm{int}}(\textbf{R}^j)}.
\label{fermionZ}
\end{equation}
In Fig.~\ref{Figure1}(a), we illustrate the case of 4 fermions across 6 imaginary time slices. The caption provides an explanation of the figure. For the one-dimensional case, it can be proven \cite{Chin1,Chin2} that \( \prod_{j=1}^M D_{\mathrm{free}}(\textbf{R}^j, \textbf{R}^{j+1}; \Delta\tau) \) is always positive, and thus the fermion sign problem does not exist. In this case, we can perform Monte Carlo importance sampling to compute the energy via the following formula:
\begin{equation}
E_F(\beta, \lambda) = -\frac{\partial \ln Z_F(\beta, \lambda)}{\partial \beta}.
\end{equation}

However, in two and three dimensions, the fermion sign problem still exists because, for  closed paths, \( \prod_{j=1}^M D_{\mathrm{free}}(\textbf{R}^j, \textbf{R}^{j+1}; \Delta\tau) \) can still be either positive or negative \cite{PB,Chin1,Chin2}. By combining a fourth-order factorization of the density matrix with fermion propagators, the permutation blocking path integral Monte Carlo (PB-PIMC) \cite{PB,PB2,PB3} developed by Dornheim \textit{et al}., has made significant contributions to the simulation of warm dense uniform electron systems by alleviating the fermion sign problem.  
However, it is still difficult to overcome the fermion sign problem in the regime of strong quantum degeneracy for large-scale fermionic systems.

\subsection{Pseudo-Fermion Propagator and Auxiliary Partition Function}

We construct the pseudo-fermion propagator \( \prod_{j=1}^M D_{\mathrm{free}}(\textbf{R}^j, \textbf{R}^{j+1}; \Delta\tau) \) by taking the absolute value of the free-fermion propagator along a closed path as follows (see the schematic in Fig.~\ref{Figure1}(b)):
\begin{equation}
D_{\mathrm{pseudo}}[\mathrm{closed~path}] = \prod_{j=1}^M \left| D_{\mathrm{free}}(\textbf{R}^j, \textbf{R}^{j+1}; \Delta\tau) \right|.
\end{equation}
This pseudo-fermion propagator is non-negative for all closed paths.

Based on the pseudo-fermion propagator, we define the following auxiliary partition function for the pseudo-fermions:
\begin{equation}
Z_{pf}(\beta, \lambda,M) = \int \prod_{j=1}^M \left| D_{\mathrm{free}}(\textbf{R}^j, \textbf{R}^{j+1}; \Delta\tau) \right|
e^{-\Delta\tau V_{\mathrm{pot}}(\textbf{R}^j)} e^{-\Delta\tau V_{\mathrm{int}}(\textbf{R}^j)}.
\end{equation}
Unlike $Z_F(\beta,\lambda)$, we will later find that $Z_{pf}(\beta,\lambda,M)$ depends on $M$. Therefore, to be safe, we have included $M$ as an argument in $Z_{pf}$. Compared to the fermionic partition function given by Eq. (\ref{fermionZ}), the pseudo-fermion partition function takes the absolute value of all \( D_{\mathrm{free}}(\textbf{R}^j, \textbf{R}^{j+1}; \Delta\tau) \). Obviously, pseudo-fermions are neither fermions nor bosons. 
For the partition function of pseudo-fermions, the previously developed techniques for handling the partition function based on the fermionic propagator \cite{Takahashi,Miura,Lyubartsev1,Lyubartsev2,Filinov1,Filinov2} can be directly applied with only minor modifications. In Appendix \ref{appendix}, we present additional details on numerical stability and the simulation methods.

\section{General Relationship Between the Energies of Pseudo-Fermions and Fermions}
\label{fdp}

\subsection{General Considerations}

The energy of the pseudo-fermions is given by:
\begin{equation}
E_{pf}(\beta, \lambda,M) = -\frac{\partial \ln Z_{pf}(\beta, \lambda,M)}{\partial \beta}.
\end{equation}
Here, \( \lambda \) represents the interaction strength between particles.

It is worth noting that $Z_F(\beta, \lambda)$  is always positive. In general, we have
\begin{equation}
Z_F(\beta, \lambda) = X(\beta, \lambda,M) Z_{pf}(\beta, \lambda,M).
\end{equation}
Due to the fermion sign problem, the factor \( X(\beta, \lambda,M) \) can be an extremely small positive number. 
We have the following expression for the energy of fermions $E_f$:
\begin{equation}
E_f(\beta, \lambda) = -\frac{\partial \ln X(\beta, \lambda,M)}{\partial \beta} + E_{pf}(\beta, \lambda,M).
\label{gr}
\end{equation}
Since $E_f$ is defined as the exact energy of fermions, it is therefore independent of $M$. In the above relation, we can obtain \( E_{pf}(\beta, \lambda,M) \) for pseudo-fermions through exact first-principles simulations free from the sign problem. However, to determine the energy of the fermionic system, we also need to know the value of  \( -\frac{\partial \ln X(\beta, \lambda,M)}{\partial \beta} \).

We define
\begin{equation}
E_X(\beta,\lambda,M)=-\frac{\partial \ln X(\beta, \lambda,M)}{\partial \beta},
\end{equation}
and thus have
\begin{equation}
E_f(\beta, \lambda) =E_X(\beta,\lambda,M)+ E_{pf}(\beta, \lambda,M).
\end{equation}
From the above exact equation, we have
\begin{equation}
\frac{\partial E_f(\beta, \lambda)}{\partial\lambda} =\frac{\partial E_X(\beta,\lambda,M)}{\partial\lambda}+ \frac{\partial E_{pf}(\beta, \lambda,M)}{\partial\lambda}.
\end{equation}

\begin{figure}[htbp]
\begin{center}
\includegraphics[width=0.7\textwidth]{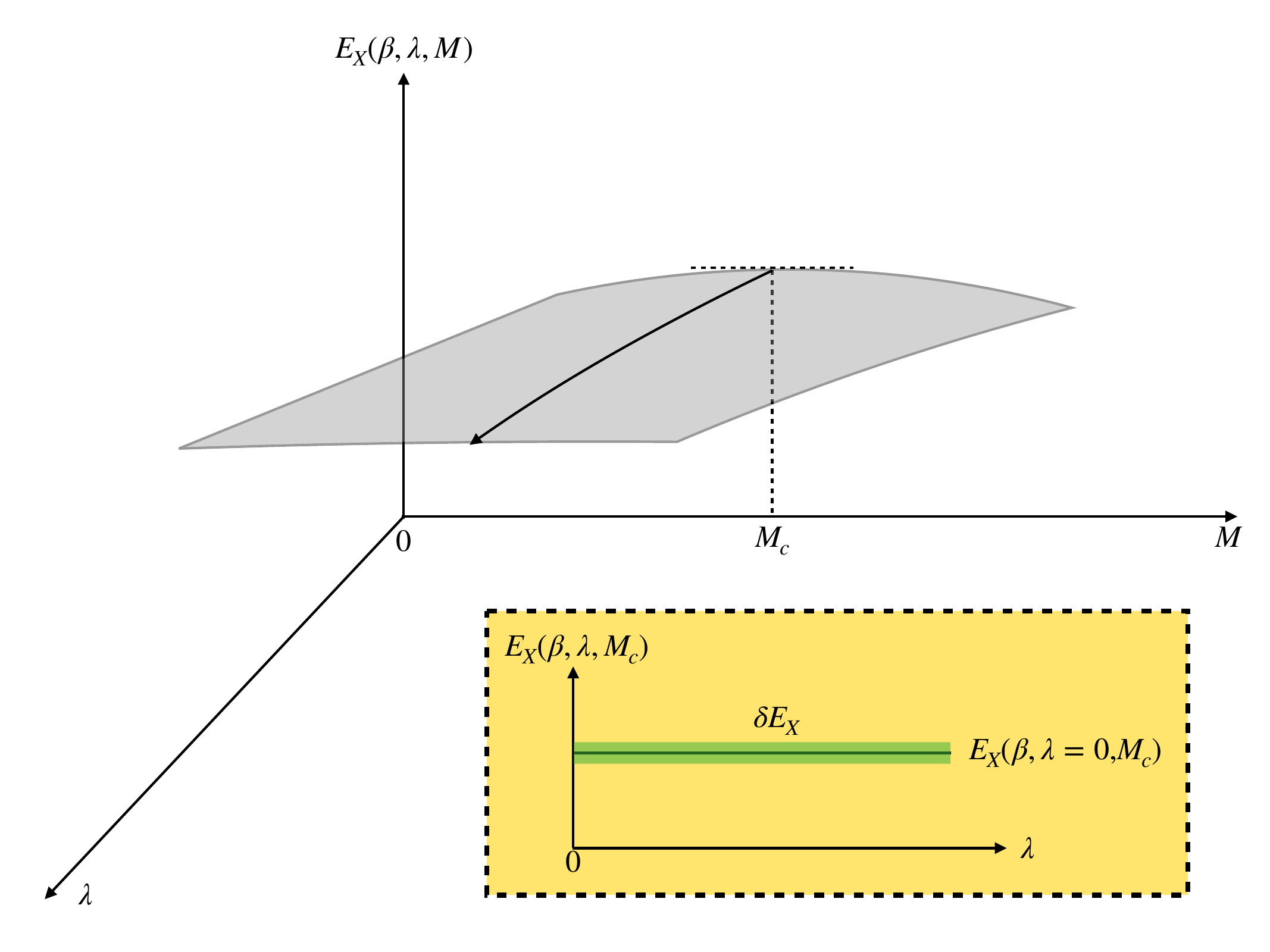}
\caption{\label{illustration} The figure shows a schematic surface plot of $E_X(\beta, \lambda, M)$. The curve $E_X(\beta, \lambda, M_c)$ originating from the point $M_c$ is indicated by an arrowed line. The horizontal dashed line represents the extremum of $E_X(\beta, \lambda=0, M)$ at $M_c$. We expect the surface to be relatively flat in the vicinity of $M_c$, so that along the arrowed curve, the relative deviation of the fermion energy inferred from first-principles simulations remains small. This is why we recommend using $M_c$ to carry out PIMC simulations in the pseudo-fermion method in this work. In the inset enclosed by the black dashed box, we show that if, in the studied $\lambda$ range, $\delta E_X \ll E_X(\beta, \lambda=0, M_c)$, the pseudo-fermion method can provide highly accurate simulations of the fermion system’s energy.}
\end{center}
\end{figure}

Although we can determine $\frac{\partial E_{pf}(\beta, \lambda, M)}{\partial \lambda}$ exactly from first-principles simulations, $\frac{\partial E_f(\beta, \lambda)}{\partial \lambda}$ is not known in advance. Therefore, what we need to do is to attempt to suppress, in a practical manner, the contribution from $\frac{\partial E_X(\beta, \lambda, M)}{\partial \lambda}$ as much as possible. Another strategy is to find a suitable fermion system for which $\frac{\partial E_X(\beta, \lambda, M)}{\partial \lambda}$ can be neglected. We note that $E_X(\beta, \lambda, M)$ represents the difference between the energy of the fermionic system and that of the pseudo-fermions. Based on this, we propose to evaluate $E_X(\beta, \lambda, M)$ at $\lambda = 0$ by performing simulations for different values of $M$. We then choose a value $M_c$ at which $E_X(\beta, \lambda=0, M)$ is minimized. In Fig. \ref{illustration}, we present a possible shape of $E_X(\beta, \lambda, M)$ and the intuition behind choosing $M_c$ for the simulations. The strategy for choosing $M_c$ is based on the conjecture that $E_X(\beta, \lambda, M)$ is also the flattest with respect to $\lambda$ at $M_c$.

If $E_X(\beta, \lambda, M_c) << E_{pf}(\beta, \lambda, M_c)$, we then expect that $\left|\frac{\partial E_X(\beta, \lambda, M_c)}{\partial \lambda}\right| << \frac{\partial E_{pf}(\beta, \lambda, M_c)}{\partial \lambda}$. Hence, we have
\begin{equation}
\frac{\partial E_f(\beta, \lambda)}{\partial\lambda} \approx \frac{\partial E_{pf}(\beta, \lambda,M_c)}{\partial\lambda}.
\end{equation}

Using
\[
E_f(\beta, \lambda) = E_f(\beta, \lambda=0) + \int_0^\lambda dx \, \frac{\partial E_f(\beta, x)}{\partial x},
\]
\[
E_{pf}(\beta, \lambda,M_c) = E_{pf}(\beta, \lambda=0,M_c) + \int_0^\lambda dx \, \frac{\partial E_{pf}(\beta, x,M_c)}{\partial x},
\]
we have
\begin{equation}
E_f(\beta, \lambda) \approx E_X(\beta, \lambda=0,M_c) + E_{pf}(\beta, \lambda,M_c).
\label{dlambda}
\end{equation}
The above equation implies that once we have determined $M_c$ and the corresponding $E_X(\beta, \lambda=0, M_c)$ in the non-interacting case, we can simulate the pseudo-fermion energy for various values of $\lambda$ and simultaneously obtain the corresponding energies of fermions. By replacing $M_c$ with other values of $M$, we can use Eq. (\ref{dlambda}) to study the inferred energy of the fermion system for different choices of $M$. We refer to the method introduced here for inferring the energy of the fermion system as the pseudo-fermion method. 

Let us consider the following form of $X(\beta, \lambda, M_c)$:
\begin{equation}
X(\beta, \lambda, M_c)=a(M_c)e^{-b(M_c)(1+c(\beta,\lambda,M_c))\beta}.
\end{equation}
Here $c(\beta,\lambda,M_c)$ is due to interparticle interaction. $c(\beta,\lambda,M_c)=0$ for $\lambda=0$. In this case, we have
\begin{equation}
E_X(\beta,\lambda,M_c)=b(M_c)+b(M_c)\frac{\partial c(\beta,\lambda,M_c)\beta}{\partial\beta}.
\end{equation}
Here $b(M_c)(\equiv E_X(\beta,\lambda=0,M_c))$ is the difference between the energy of fermions and pseudo-fermions at $\lambda=0$, which can be considered in an exact way. The second term on the right-hand side of the above equation is the deviation that the current pseudo-fermion method cannot incorporate. Since this deviation has no contribution at $\lambda = 0$, we have $\frac{\partial c(\beta, \lambda=0, M_c)\beta}{\partial \beta} = 0$. Therefore, at least for relatively small $\lambda$, we may omit $\frac{\partial c(\beta, \lambda=0, M_c)\beta}{\partial \beta}$. The ratio of the second term on the right-hand side of the above equation to the energy of the non-interacting fermion system is given by $\frac{b(M_c)}{E_f(\lambda=0)} \frac{\partial c(\beta, \lambda, M_c)\beta}{\partial \beta}$. In first-principles simulations, we choose the optimal $M_c$ to make $b(M_c)/E_f(\lambda=0)$ as small as possible, and we know the minimum value of this quantity. It appears that by minimizing $\frac{b(M_c)}{E_f(\lambda=0)}$, we may simultaneously suppress $\frac{\partial c(\beta, \lambda, M_c)}{\partial \beta}$ as much as possible. 

\subsection{Evaluation of the relative deviation in the pseudo-fermion method}

We provide a heuristic discussion on evaluating the relative deviation caused by the pseudo-fermion method. As illustrated in the inset enclosed by the black dashed
box in Fig. \ref{illustration}, we define $\delta E_X$ as the maximum deviation in the studied $\lambda$ range compared to $\lambda = 0$. The relative deviation of the fermion system’s energy simulated using the pseudo-fermion method is
\begin{equation}
\Delta_f(\lambda) \sim \frac{E_X(\beta, \lambda=0, M_c)}{E_f(\beta, \lambda=0)} \times \frac{\delta E_X}{E_X(\beta, \lambda=0, M_c)}.
\end{equation}
When $\delta E_X << E_X(\beta, \lambda=0,M_c)$, $\Delta_f(\lambda)$ is the product of two small quantities; therefore, in this case, the pseudo-fermion method can provide highly accurate simulations of the fermion system’s energy. For example, if for a certain fermion system we find that $E_X(\beta,\lambda=0,M_c)/E_f(\lambda=0) = 1\%$, then we can expect the relative deviation caused by pseudo-fermion method to be much below $1\%$ even in the presence of interactions if the surface illustrated in Fig. \ref{illustration} does not vary drastically near $M_c$. 
In our later simulation examples (Sec. \ref{results}), by comparing with existing benchmarks for quantum dots, we find that even for large $\lambda$, the relative deviation caused by the pseudo-fermion method can be an order of magnitude smaller than $E_X(\beta,\lambda,M_c)/E_f(\lambda=0)$. The underlying reason is that the variation amplitude of the curve $E_X(\beta, \lambda, M_c)$ with respect to $\lambda$ is much smaller than $E_X(\beta, \lambda=0, M_c)$.


\subsection{General procedure of the pseudo-fermion method}

We summarize here the procedure for inferring the energies of fermions using the pseudo-fermion method:

\begin{enumerate}

\item For $\lambda = 0$, we simulate $E_{pf}(\beta, \lambda=0, M)$ for different values of $M$. Additionally, we calculate the energy of the fermion system $E_f(\beta, \lambda)$ at $\lambda = 0$ (\textit{e.g.}, recurrence relations\cite{Borrmann} in the canonical ensemble or grand canonical ensemble when it is appropriate). Then, we determine $M_c$ such that $E_{X}(\beta, \lambda=0, M)$ is minimized.

\item	Compute the energy difference $ E_X(\beta,\lambda=0,M_c) = E_f(\beta,\lambda=0) - E_{pf}(\beta,\lambda=0,M_c)$;

\item	For $\lambda \neq 0$, perform a PIMC simulation to obtain the energy $E_{pf}(\beta,\lambda,M_c)$ of pseudo-fermions;

\item	Infer the energy of fermions as $E_f(\beta,\lambda) = E_{X}(\beta,\lambda=0,M_c) + E_{pf}(\beta,\lambda,M_c)$.

\end{enumerate}

\section{Results}
\label{results}

Obviously, any method that aims to overcome the fermion sign problem only demonstrates its value when interactions between particles are present. In this work, we consider \(N\) spin-polarized fermions in a quantum dot \cite{Reimann,Waltersson,WangL}. The dimensionless Hamiltonian is (\(\hbar = \omega = k_B = m = 1\)):
\begin{equation}
\hat{H} = -\frac{1}{2} \sum_{i=1}^N \Delta_i + \frac{1}{2} \sum_{i=1}^N \mathbf{r}_i^2 + \sum_{i<j} \frac{\lambda}{|\mathbf{r}_i - \mathbf{r}_j|}.
\end{equation}

We will use the pseudo-fermion method to infer the energy of fermions in quantum dots, considering various situations ranging from the ground state and strong quantum degeneracy to weak quantum degeneracy. 
In this work, we use $2 \times 10^7$ Monte Carlo steps for thermalization and $2 \times 10^8$ Monte Carlo steps for energy sampling. For each set of parameters, we performed ten independent simulations and calculated the mean and standard deviation. 

\subsection{Simulation of approximately ground states}

We first consider the case of $N=8,\beta=10$. For this situation, which can be approximated as the ground state, existing benchmarks\cite{Egger} are available for comparison because the number of particles is small.

\begin{figure}[htbp]
\begin{center}
\includegraphics[width=0.8\textwidth]{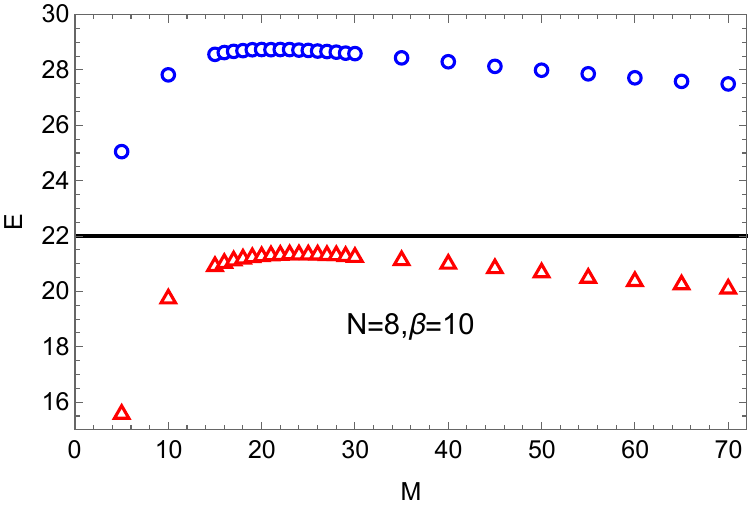} 
\caption{\label{N8beta10M} For $N = 8, \beta = 10$ and different numbers of imaginary time slices, $M$, the red triangles and blue circles represent the energies of the pseudo-fermions at $\lambda=0$ and $\lambda=0.5$, respectively. The black line represents the exact energy of the non-interacting fermions.}
\end{center}
\end{figure}

In Fig. \ref{N8beta10M}, the red triangles represent the pseudo-fermion energy for the non-interacting case ($\lambda=0$) at various imaginary time slices, $M$. The black line represents the exact energy of the non-interacting fermions. We observe the following:

\begin{enumerate}

\item $E_{pf} (\lambda=0,M)$ depends on $M$ and does not converge as $M$ increases. This provides evidence that $Z_{pf}$ depends on $M$.

\item For all values of $M$ shown, $E_{pf} (\lambda=0,M)$ is consistently less than the actual energy of non-interacting fermions.

\item At $M_c=23$, the pseudo-fermion energy is closest to the energy of fermions, and the energy changes slowly in the vicinity of $M_c=23$.

\end{enumerate}

Based on the analysis in the previous section (Sec. \ref{fdp}), we may consider selecting $M_c=23$ when using the pseudo-fermion method to infer the energy of fermions. We note that this method for choosing $M_c$ is unique, which makes it both practical and independently verifiable by other methods. For $M_c = 23$, $E_X(\lambda=0) / E_{f}(\lambda=0) \approx 0.03$, so we expect the pseudo-fermion method to accurately estimate the energy of fermions. In Fig. \ref{N8beta10M}, we use blue circles to show the pseudo-fermion energy for $\lambda=0.5$. We observe that the energy of the interacting pseudo-fermions varies with $M$ in a similar way to the $\lambda=0$ case. 

\begin{figure}[htbp]
\begin{center}
\includegraphics[width=0.8\textwidth]{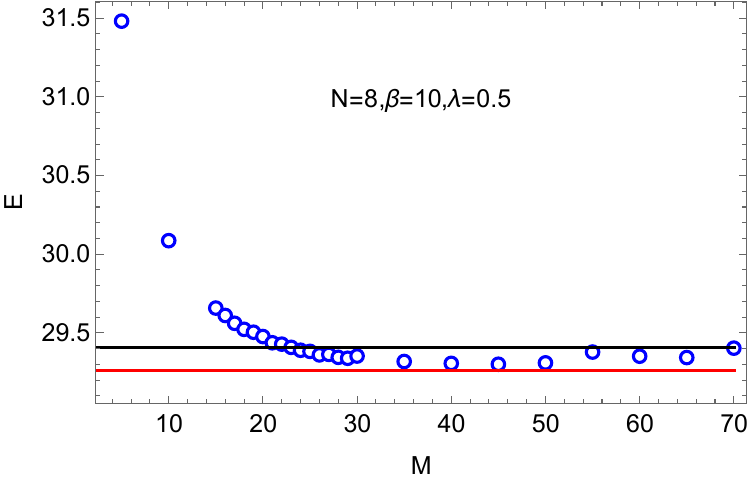} 
\caption{\label{N8converge} For $N=8,\beta=10$, the blue circles represent the inferred energies of fermions at different values of $M$, calculated using the pseudo-fermion method. The black line represents the inferred fermion energy at $M=23$, and the red line represents the energy after shifting the black line downward by $0.5\%$.}
\end{center}
\end{figure}

Of course, we can actually infer the energy of the fermion system based on the pseudo-fermion energy simulation for any value of $M$. For the case $M \neq M_c$, as long as $E_X(\beta, \lambda, M)$ depends weakly on $\lambda$, we can still reliably infer the energy of the fermion system. In Fig. \ref{N8converge}, the blue circles represent the inferred fermion energy as a function of $M$. The black line represents the inferred fermion energy at $M=23$, and the red line represents the energy after shifting the black line downward by $0.5\%$. We note that in the range $23\leq M\leq 70$, the inferred fermion energy only varies by less than $0.5\%$. This result implies that throughout the entire range $23 \leq M \leq 70$, $\frac{\partial E_X(\beta, \lambda, M)}{\partial \lambda} \approx 0$. When $\left| \frac{E_X(\beta, \lambda=0)}{E_{f}(\beta, \lambda=0, M)} \right| << 1$, the only reason for the possible failure of the pseudo-fermion method comes from a very large $\frac{\partial E_X(\beta, \lambda, M)}{\partial \lambda}$. 

\begin{figure}[htbp]
\begin{center}
\includegraphics[width=0.8\textwidth]{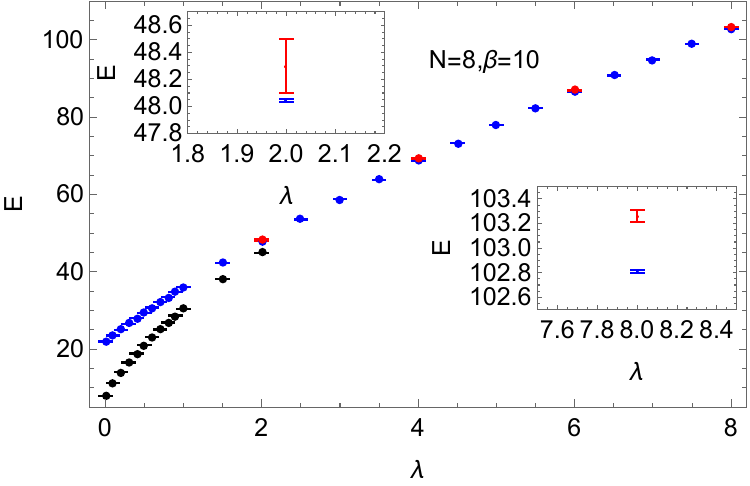} 
\caption{\label{N8beta10} For the case of $N=8, \beta=10$, the blue dots and the red dots with error bars represent the simulation results obtained using the pseudo-fermion method and the MLB algorithm \cite{Egger}, respectively. In the inset, we present a magnified view of the simulation results for $\lambda = 2$ and $\lambda = 8$. The black dots with error bars represent the energy of bosons simulated using the worm algorithm.}
\end{center}
\end{figure}

For the case of $N = 8, \beta = 10$, Egger \textit{et al.}\cite{Egger} provided the energies of fermions based on the multilevel blocking (MLB) algorithm. As shown in Fig. \ref{N8beta10}, we find excellent agreement between the results of the pseudo-fermion method (blue dots with error bar) and the MLB algorithm (red dots with error bar). In the inset of Fig. \ref{N8beta10}, we present a magnified view of the simulation results for $\lambda = 2$ and $\lambda = 8$. Throughout the entire $\lambda$ range, the deviation from the MLB results does not exceed $0.5\%$, which is an order of magnitude smaller than $E_X(\lambda=0)/E_{f}(\lambda=0)$ at $M_c$. We also observe that the relationship between energy \(E_f(\lambda)\) and \(\lambda\) is not linear, but the pseudo-fermion method can still perfectly capture the nonlinear behavior. As $\lambda$ decreases, the sign problem becomes increasingly severe in the MLB algorithm~\cite{Egger}, leading to larger fluctuations due to the sign factor. In contrast, our simulations are not affected by this factor, allowing for reliable simulations in the small-$\lambda$ regime.

Especially for large $\lambda$, the pseudo-fermion method is always biased. Since a large repulsive interaction $\lambda$ suppresses the fermion sign problem, the MLB \cite{Egger} and PB-PIMC (permutation blocking path integral Monte Carlo) \cite{PB} methods can provide highly accurate simulations. A comparison with the results from the pseudo-fermion method allows us to estimate the bias introduced by $\lambda$. Furthermore, because the pseudo-fermion method becomes more reliable as $\lambda$ decreases, we can establish a clear upper bound for the bias at small $\lambda$ based on the basis analysis at large $\lambda$.  
Since the upper bound for the bias can be reasonably estimated, we believe that our results in the small-$\lambda$ regime can serve as valuable benchmarks.

For comparison, in Fig. \ref{N8beta10}, the black dots with error bars represent the energy of bosons simulated using the worm algorithm \cite{Burov1,Burov2,Spada,Tommaso}. We observe that the boson energy is lower than the energy of fermions, and its dependence on $\lambda$ is very different. This makes it impossible to accurately infer the energies of fermions through a simple energy shift for bosons, unlike the pseudo-fermion approach. Moreover, we find that as $\lambda$ increases, the energy of fermions changes more gradually than that of bosons. This is a characteristic feature of Fermi statistics. Due to the Pauli exclusion principle, two fermions cannot come arbitrarily close to each other, which suppresses the effect of interactions compared to bosons. 

\subsection{Simulation of strong quantum degeneracy}

\begin{figure}[htbp]
\begin{center}
 \includegraphics[width=0.8\textwidth]{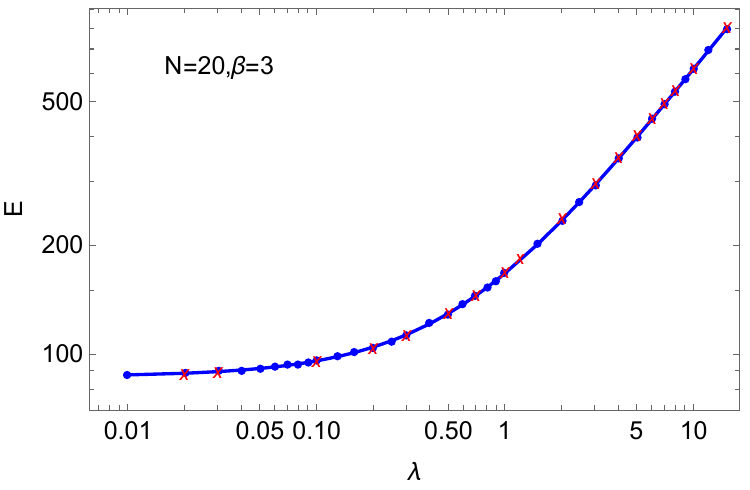} 
\caption{\label{N20beta3} For \(N=20, \beta=3\), the blue dots represent the inferred fermion system energy based on pseudo-fermion method,  connected by a blue line. The red crosses are energy results from Ref. \cite{PB} based on CPIMC and PB-PIMC.}
\end{center}
\end{figure}

The pseudo-fermion method allows us to simulate the thermodynamic properties of large-scale fermion systems. In Fig.~\ref{N20beta3}, we present simulation results for \(N=20\) at \(\beta=3\). At this temperature $T = 1/3$, the system is in the regime of strong quantum degeneracy (the Fermi degeneracy energy is 6). Meanwhile, for the case of $N = 20$, the usual direct PIMC \cite{Dornheim} fails to provide reliable results at small $\lambda$ due to the severe fermion sign problem\cite{PB}. 

In this example, based on the relationship between the energy of the non-interacting pseudo-fermions and $M$, $M_c$ is chosen to be $9$. For this choice of $M_c$, $E_X(\lambda=0) / E_{f}(\lambda=0) \approx 0.015$.
In Fig.~\ref{N20beta3}, the blue dots represent the inferred energy of the fermionic systems based on pseudo-fermion method, while the red crosses are the results of CPIMC (configuration path integral Monte Carlo) and PB-PIMC \cite{PB} by Dornheim \textit{et al.}. Across all different values of $\lambda$, the pseudo-fermion method shows excellent agreement with the CPIMC and PB-PIMC results. 
 
In PB-PIMC \cite{PB}, the average sign is smaller than $10^{-3}$ for $\lambda < 1$, making accurate simulations extremely time-consuming, while it is difficult for CPIMC to simulate fermions for $\lambda > 0.3$.  
In contrast, the pseudo-fermion approach is much more efficient in the whole regime because there is no sign problem during the simulation. This speedup becomes even more significant with larger particle numbers or lower temperatures. Owing to the high efficiency of the pseudo-fermion method, we present in Fig.~\ref{N20beta3} a large number of simulation results as blue dots, connected by a blue line.
Since the fluctuations in the simulations are below $0.01\%$, the error bars are omitted from the figure for clarity. 

In this example, the energy of fermions can already be accurately simulated at $M_c=9$, whereas a larger $M$ is required to assure convergence when we simulate boltzmannons and bosons. One of the reasons for the faster convergence of the energy of fermions in our method is that the determinant between neighboring imaginary time slices has already accurately accounted for the fermionic exchange antisymmetry. As a result, fermions are prevented from coming too close to each other, effectively suppressing the short-range repulsive interaction.

\subsection{Quantum dots with $\beta=10$ and $N=\{20,50\}$}
\label{Mchoice}

\begin{figure}[htbp]
\begin{center}
 \includegraphics[width=0.8\textwidth]{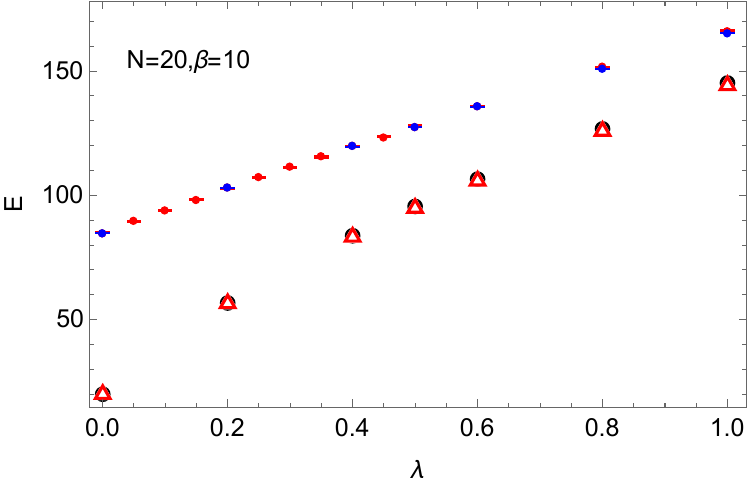} 
\caption{\label{N20beta10} For the case of $N=20$ and $\beta=10$, the blue dots and the red dots represent the fermionic energy obtained from the pseudo-fermion method for $M=32$ and $M=100$, respectively. The black circles and red triangles denote the energies of boltzmannons and bosons, respectively.}
\end{center}
\end{figure}

We now consider the case of $N=20$ and $\beta=10$. In this example, the fermion sign problem is so severe that neither the PB-PIMC approach\cite{PB} nor the MLB\cite{Egger} algorithm can reliably perform simulations for small $\lambda$. Although the temperature is the same as in the previous $N=8$ example, the larger number of particles in the current case means the quantum degeneracy effect is stronger. Therefore, we need to re-decide the choice of $M$ using the method introduced earlier. For this case, we recommend $M=32$, and $E_X(\lambda=0) / E_{f}(\lambda=0) \approx 0.007$.
In Fig. \ref{N20beta10}, we present the simulation results (blue dots with error bars) obtained using the pseudo-fermion method with $M=32$, and the red dots with error bars represent the simulation results with $M = 100$. We note that the fermionic energies obtained with $M = 32$ and $M = 100$ are in excellent agreement. This comparison shows that even though we recommend $M=32$, the energy of fermions is not sensitive to the choice of $M$. This means that $E_X(\beta, \lambda, M)$ is flat over a wide range of $M$. When no reference benchmark is available, checking the magnitude of $E_X(\lambda=0) / E_{f}(\lambda=0)$ and the correlation between the inferred fermion system energy and different $M$ can provide a way to assess the reliability of the pseudo-fermion method and the size of its deviation. The smaller $E_X(\lambda=0) / E_{f}(\lambda=0)$ is and the weaker the dependence of the inferred fermion energy on $M$, the more likely the pseudo-fermion method is to be accurate.

The black circles and red triangles in Fig. \ref{N20beta10} represent the energies of boltzmannons and bosons, respectively, obtained using the worm algorithm \cite{Burov1,Burov2,Spada,Tommaso}. 
For bosons and boltzmannons, we chose $M = 100$ to ensure the convergence of the simulation.
Since bosons and boltzmannons share the same energy in the ground state, the blue circles and red triangles essentially overlap. We immediately observe that shifting the energy of bosons or boltzmannons cannot reproduce the energy of fermions. 

In Table~\ref{N50beta10}, we present the energy simulation results for $N = 50, \beta = 10$ based on the pseudo-fermion method. For this case, we recommend $M=38$, and $E_X(\lambda=0) / E_{f}(\lambda=0) \approx 0.001$. In this case, we have good opportunity to keep the deviation below $0.1\%$ in the small $\lambda$ region. We believe the results in Table~\ref{N50beta10} will be helpful for future verification through other methods and may serve as benchmarks for developing new approaches. 
Incidentally, due to the fermion sign problem, reliable PIMC benchmarks for more than $50$ particles in two-dimensional quantum dots at low temperatures and small $\lambda$ are still highly lacking.
\begin{table}
    \centering
    \begin{tabular}{cc}
    $\lambda$  & $E_f(\lambda)$ \\ 
        0.0 & 335.00 $\pm$ 0.037 \\
        0.01 & 340.01 $\pm$ 0.08 \\
        0.02 & 345.01 $\pm$ 0.09 \\
        0.03 & 349.99 $\pm$ 0.07 \\
        0.04 & 354.94 $\pm$ 0.07 \\
        0.05 & 359.84 $\pm$ 0.06 \\
        0.06 & 364.74 $\pm$ 0.07 \\
        0.07 & 369.63 $\pm$ 0.08 \\
        0.08 & 374.49 $\pm$ 0.08 \\
        0.09 & 379.35 $\pm$ 0.08 \\
        0.1 & 384.16 $\pm$ 0.07 \\
        0.2 & 431.40 $\pm$ 0.07 \\ 
       0.4  & 520.96 $\pm$ 0.09 \\ 
       0.5  & 563.71 $\pm$ 0.07 \\ 
       0.6  & 605.23 $\pm$ 0.07 \\
       0.8  & 685.15 $\pm$ 0.08 \\
        1.0 & 761.50 $\pm$ 0.07 \\
    \end{tabular}
    \caption{The energy of $50$ fermions with $\beta=10$ in quantum dots obtained using the pseudo-fermion method.}
    \label{N50beta10}
\end{table}

\subsection{An Application Covering Simulations from Weak to Strong Quantum Degeneracy}

Let us now consider a particularly valuable example: $N=6, \beta=1$ and $0 \leq \lambda \leq 20$.
For such a small particle number at this temperature, Dornheim \textit{et al.} \cite{Dornheim1} obtained highly accurate fermion energies using the direct PIMC approach for $0\leq\lambda\leq 1$.
As $\lambda$ increases, the fermionic system undergoes a transition from strong quantum degeneracy to weak quantum degeneracy.
At $\lambda = 1$, which corresponds to weak quantum degeneracy, Dornheim \textit{et al.} \cite{Dornheim1} found that the isothermal $\xi$-extrapolation method based on fictitious identical particles can reliably infer the fermion energy.
However, for $\lambda = 0$ and $0.2$, the isothermal $\xi$-extrapolation method completely fails.
We now investigate whether the pseudo-fermion method can provide reliable results over the entire range $0 \leq \lambda \leq 20$.

\begin{figure}[htbp]
\begin{center}
\includegraphics[width=0.8\textwidth]{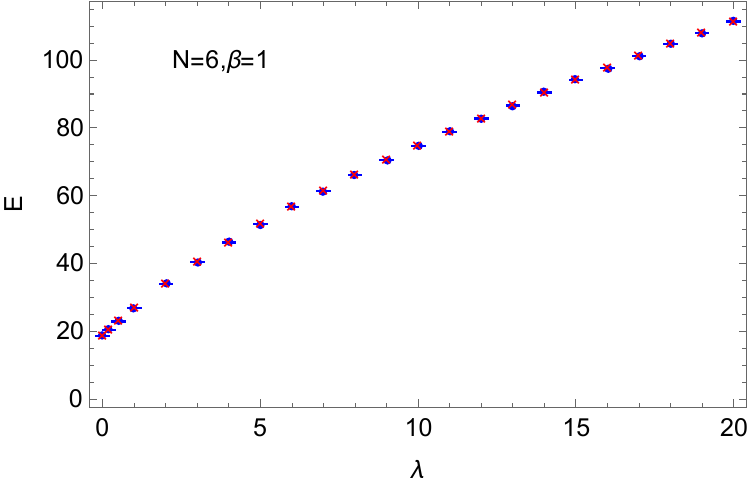} 
\caption{\label{N6beta1} For $N=6,\beta=1$, the energies based on the pseudo-fermion method are shown as blue dots with error bars, while the red crosses represent the direct PIMC results.}
\end{center}
\end{figure}

Starting with simulations of non-interacting pseudo-fermions for various values of $M$ and comparing the energy of non-interacting fermions, we find that $M_c=3$, an exceptionally small number of imaginary time slices, yields the closest agreement between the non-interacting pseudo-fermions and fermions. For this choice of $M$, $E_X(\lambda=0) / E_{f}(\lambda=0) \approx 0.02$.
We then applied the pseudo-fermion method to infer the fermion energies over the entire range $0 \leq \lambda \leq 20$.
In Fig.~\ref{N6beta1}, the results from the pseudo-fermion method are shown as blue dots with error bars, while the red crosses represent the direct PIMC results with worm algorithm in this work. We find that the fermionic energies obtained from direct PIMC and the pseudo-fermion method are in excellent agreement across the entire $\lambda$ range.

\begin{figure}[htbp]
\begin{center}
\includegraphics[width=0.8\textwidth]{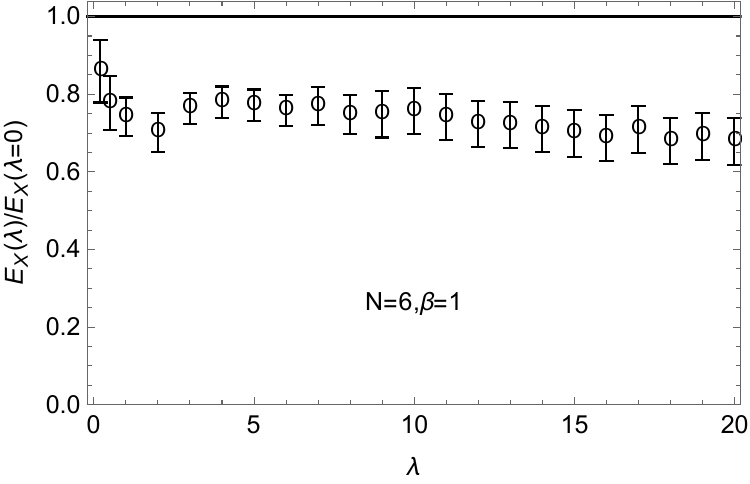} 
\caption{\label{N6beta1Ex} For $N=6,\beta=1$, the black circles with error bars represent $E_X(\lambda) / E_X(\lambda=0)$, while he horizontal black line of height $1$ serves as a reference to indicate the variation range of $E_X(\lambda) / E_X(\lambda=0)$.}
\end{center}
\end{figure}

For the case of $N = 6, \beta = 1$, since we can use direct PIMC to exactly simulate all situations with $\lambda \geq 0$, this provides an opportunity to examine the behavior of $E_X(\lambda)=E_f(\lambda)-E_{pf}(\lambda)$. In Fig.~\ref{N6beta1Ex}, the black circles with error bars show $E_X(\lambda)/E_X(\lambda=0)$ for different $\lambda$. The horizontal black line of height $1$ serves as a reference to indicate the variation range of $E_X(\lambda) / E_X(\lambda=0)$. We immediately notice that the variation amplitude of $E_X(\lambda)$ over the entire range of $\lambda$ shown is much smaller than $E_X(\lambda = 0)$. It is precisely because $E_X(\lambda=0) / E_{f}(\lambda=0) << 1$ and $\big(E_X(\lambda) - E_X(\lambda=0)\big) / E_X(\lambda=0) << 1$ that the pseudo-fermion method can infer the fermionic energy in this example with very high accuracy, \textit{i.e.}, with only a very small relative deviation. For instance, the relative deviation is only $0.1\%$ at $\lambda = 20$ and $0.4\%$ at $\lambda = 1$.

In summary, the examples presented in this section—ranging from the ground state to strong and weak quantum degeneracy—demonstrate that the pseudo-fermion method holds promise for efficient and reliable simulations across all these regimes.
Since the pseudo-fermion method is free from the sign problem during simulations, it shows considerable potential for future practical applications.

\section{Summary and Discussion}
\label{summary}

In summary, in this work we have found that pseudo-fermions provide a new possibility for simulating fermionic systems. Based on comparisons with existing benchmarks\cite{PB, Egger,Dornheim1} for simulations of quantum dots in two-dimensional harmonic traps, we found that the pseudo-fermion method can efficiently produce reliable simulation results for situations with different degrees of quantum degeneracy. One key reason for the reliable simulation of the energies of fermionic systems by the pseudo-fermion method lies in the fact that the antisymmetric exchange of fermions has already been fully incorporated in the pseudo-fermion propagators between neighboring imaginary time slices. Pseudo-fermions are neither bosons nor fermions, but when analyzing only neighboring imaginary time slices, pseudo-fermions satisfy the Pauli exclusion principle of fermions. Therefore, pseudo-fermions are closer in nature to fermions. Another key reason for the success of the pseudo-fermion method is the flat behavior of $E_X(\beta, \lambda, M)$ near $M_c$, as illustrated in Fig. \ref{illustration} and confirmed by the examples in this paper.

In the simulation techniques using fictitious identical particles \cite{XiongFSP,Xiong-xi} to overcome the fermion sign problem, the real parameter $\xi$ associated with the fictitious identical particles can be easily incorporated \cite{Dornheim1, Morresi1} into the widely used worm algorithm \cite{Burov1,Burov2,Spada}. Similar to the fictitious identical particles, the simulation of pseudo-fermions in this work only requires minor technical modifications to the existing technique \cite{Lyubartsev1} of fermionic propagator PIMC. In the fixed-node method \cite{fnm} and the restricted path integral Monte Carlo \cite{RPIMC}, one needs to incorporate trial wavefunction fixed nodes and restricted regions, whereas the pseudo-fermion method is simpler, more direct, and offers easier extensibility. For example, the pseudo-fermion method handles cases near zero temperature and at finite temperature in exactly the same way.

The purpose of this paper is to establish a method for simulating fermionic systems based on pseudo-fermions, rather than to focus on applications to specific real systems. However, we believe that pseudo-fermions have potential applications in many areas involving fermionic physics, such as warm dense matter \cite{WDM,Bonitz-Review,RoadMap} and the Fermi-Hubbard model \cite{LeBlanc,Arovas}. 
For weakly interacting fermionic systems, our analysis indicates that the pseudo-fermion method is most likely to yield highly accurate results, making it a promising approach for first-principles simulations of ultracold Fermi gases \cite{lattice,Bloch,Fermilattice}. In studies of quantum phase transitions \cite{Vojta}, one often needs to investigate the energy changes as the interaction strength varies at a given low temperature. We expect that pseudo-fermions will prove valuable for such studies in the future, because in these cases it is not necessary to know the exact energy at the critical interaction strength \(\lambda_c\); it suffices to simulate the change in energy near \(\lambda_c\). As shown in Fig. \ref{illustration}, if $E_X(\beta, \lambda, M_c)$ changes slowly near $\lambda_c$, we can accurately model the relationship between energy and $\lambda$ near $\lambda_c$. For weakly or moderately quantum-degenerate systems like warm dense matter, the isothermal $\xi$-extrapolation method \cite{XiongFSP, Dornheim1} is a more suitable approach at the current stage than the pseudo-fermion method, especially since Dornheim \textit{et al.}, found that the isothermal $\xi$-extrapolation method can simulate a very rich variety of thermodynamic properties of fermions in a series of groundbreaking works \cite{Dornheim1,Dornheim2,Dornheim3,Dornheim4,Dornheim5,Dornheim6,Dornheim7}. Applying the pseudo-fermion method to thermodynamic properties other than energy is worth pursuing in future research. For strongly quantum degeneracy or zero temperature, the isothermal $\xi$-extrapolation method can not \cite{XiongFSP,Dornheim1,He} give reliable simulation of fermions. In a sense, pseudo fermions and fictitious identical particles\cite{XiongFSP,Xiong-xi} are complementary to addressing the problem of fermion sign problem.

\begin{acknowledgments}
This work has received funding from Hubei Polytechnic University. We thank Dr. Tommaso Morresi for reading the draft and providing valuable suggestions for improvement.
\end{acknowledgments}






\appendix
\renewcommand{\thefigure}{A\arabic{figure}}
\setcounter{figure}{0}

\section{Numerical stability and simulation method}
\label{appendix}

\subsection{Numerical stability}
Since each element of the matrix involved in the determinant $D_{free}(\mathbf{R}^j,\mathbf{R}^{j+1};\Delta\tau)$ is an exponential function, we need to take care in its numerical evaluation especially in the small $\Delta\tau$ case. To ensure numerical stability, first we calculate a value $\tilde p$ as the maximum of all the exponents in a given matrix:
\begin{equation}
\tilde p=\max_{l,m}[-\frac{1}{2\Delta\tau}(\mathbf{R}^j_l-\mathbf{R}^{j+1}_m)^2].
\end{equation}
Then the determinant can be expressed in a form that is numerically stable:
\begin{equation}
\det(\exp(-\frac{1}{2\Delta\tau}(\mathbf{R}^j_l-\mathbf{R}^{j+1}_m)^2))=\exp(N\tilde p)\times\det(\exp(-\frac{1}{2\Delta\tau}(\mathbf{R}^j_l-\mathbf{R}^{j+1}_m)^2-\tilde p)),
\end{equation}
since the maximum element of the new matrix is always 1.
\subsection{Simulation method}
In order to perform PIMC simulation for pseudo-fermions, we use standard Metropolis procedure with some Monte Carlo moves. In this work we follow the approach in Ref. \cite{Lyubartsev1}, where we have two types of random uniform moves.
\par
1. First we randomly select an imaginary time slice with index $j$, then we move all the particles in that time slice by different random uniform vectors $\Delta\mathbf{r}^j_l$ in the interval $[-r_{max}/2,r_{max}/2]$, namely
\begin{equation}
\mathbf{R}'^j_l=\mathbf{R}^j_l+\Delta\mathbf{r}^j_l.
\end{equation}
Since only the propagators for the imaginary time slices adjacent to time slice $j$ are affected, the acceptance probability for this move is just
\[
A(\mathbf{R}',\mathbf{R})=
\]
\begin{equation}
\min\left\{1,\left|\frac{D_{free}(\mathbf{R}'^{j-1},\mathbf{R}'^{j};\Delta\tau)D_{free}(\mathbf{R}'^j,\mathbf{R}'^{j+1};\Delta\tau)}{D_{free}(\mathbf{R}^{j-1},\mathbf{R}^{j};\Delta\tau)D_{free}(\mathbf{R}^j,\mathbf{R}^{j+1};\Delta\tau)}\right|\exp[-\Delta\tau(V(\mathbf{R}'^j)-V(\mathbf{R}^j))]\right\},
\end{equation}
where if $j-1<1$, then ${R}^{j-1}={R}^{j-1+M}$, and if $j+1>M$, then ${R}^{j+1}={R}^{j+1-M}$. $V$ is the total interaction potential.
\par
2. We randomly select an imaginary time slice $j$, and move all particles in the three adjacent time slices with indices $j-1$, $j$, and $j+1$ by the same random vector $\Delta\mathbf{r}$ from the uniform distribution $[-r_{max}/2,r_{max}/2]$. That is,
\begin{equation}
\mathbf{R}'^k_l=\mathbf{R}^k_l+\Delta\mathbf{r},
\end{equation}
for $k=j-1,j,j+1$. Again only the adjacent propagators are affected. The acceptance probability for this move is given by
\[
A(\mathbf{R}',\mathbf{R})=
\]
\begin{equation}
\min\left\{1,\left|\frac{D_{free}(\mathbf{R}'^{j-2},\mathbf{R}'^{j-1};\Delta\tau)D_{free}(\mathbf{R}'^{j+1},\mathbf{R}'^{j+2};\Delta\tau)}{D_{free}(\mathbf{R}^{j-2},\mathbf{R}^{j-1};\Delta\tau)D_{free}(\mathbf{R}^{j+1},\mathbf{R}^{j+2};\Delta\tau)}\right|\exp[-\Delta\tau\sum_{k=j-1}^{j+1}(V(\mathbf{R}'^k)-V(\mathbf{R}^k))]\right\}.
\end{equation}
Similarly if $j-2<1$, then ${R}^{j-2}={R}^{j-2+M}$, and if $j+2>M$, then ${R}^{j+2}={R}^{j+2-M}$.
\par
In this work we also use the thermodynamic energy estimator
\begin{equation}
E=-\frac{\partial\ln Z_{pf}}{\partial\beta}.
\end{equation}
After some calculations, it can be shown that the thermodynamic energy estimator is given by the following formula
\begin{equation}
E=\frac{dN}{2\Delta\tau}+\sum_{j=1}^M\left[\langle -Tr(A(j,j+1)^{-1}\frac{\partial}{\partial\beta}A(j,j+1))\rangle+\langle \frac{V(\mathbf{R}^j)}{M}\rangle\right],
\end{equation}
where $A(j,j+1)$ is a matrix with elements
\begin{equation}
A(j,j+1)_{l,m}=\exp(-\frac{1}{2\Delta\tau}(\mathbf{R}^j_l-\mathbf{R}^{j+1}_m)^2).
\end{equation}

\end{document}